\documentclass[preprint]{revtex4}
\usepackage{graphicx}
\newcommand{\be}{\begin{equation}}
\newcommand{\ee}{\end{equation}}
\begin{document}
\title{Quasi-Adiabatic Continuation in Gapped Spin and Fermion Systems:
Goldstone's Theorem and Flux Periodicity}
\author{M.~B.~Hastings}
\affiliation{Center for Nonlinear Studies and Theoretical Division,
Los Alamos National Laboratory, Los Alamos, NM, 87545\\
hastings@lanl.gov}
\begin{abstract}
We apply the technique of quasi-adiabatic continuation to study
systems with continuous symmetries.
We first derive a general form
of Goldstone's theorem applicable to gapped nonrelativistic systems
with continuous symmetries.  We then show that
for a fermionic system
with a spin gap, it is possible to insert $\pi$-flux into a cylinder
with only exponentially 
small change in the energy of the system, a scenario which covers
several physically interesting
cases such as an $s$-wave superconductor or a resonating valence bond
state.
\end{abstract}
\maketitle

Systems with a spin gap are believed to have a number of
interesting properties.  One expects these systems to have
short-range spin correlations, but in return exotic properties such
as fractionalization and topological order often appear\cite{general}.
In this paper, we apply the recently developed technique of quasi-adiabatic
continuation\cite{zl} to study these systems, generalizing ideas developed
in \cite{lsmhd}.  We start by
proving a version of Goldstone's theorem, demonstrating exponential decay of
spin correlations in a gapped system with continuous symmetries.  This result is
stronger than previous results\cite{loc,hk,ns} which proved exponential decay
of {\it connected} correlation functions in gapped systems, as in this
case we also show that for a system with multiple ground states the
average of the spin correlations over different ground states is exponentially
small.

We then consider the problem of
flux quantization in electron systems with a spin gap (without requiring that
there be any gap to spinless excitations).  We consider
a system of spinful fermions
which is periodic in at least one direction,
and show that if the system has a spin gap 
then there is a state which has an expectation value of the energy for
a Hamiltonian with $\pi$ flux which is close to the ground state energy of
the Hamiltonian with no flux\cite{nbp}.  This result covers various
interesting physical cases such as superconductors\cite{by} and resonating
valence bond states\cite{e2e}.

Throughout, we consider {\it local} Hamiltonians on a lattice as
defined below.  We begin by defining some notation and defining conditions
for a Hamiltonian to be local.  Then we review the technique of quasi-adiabatic
continuation, and give a brief discussion of Berry's phase for
this continuation.
We then consider Goldstone's theorem and the flux periodicity
problem.

We use labels $i,j,...$ to denote lattice sites, and introduce a metric
${\rm dist}(i,j)$.  We assume that there is a finite dimensional Hilbert
space on each site.  An example of a local Hamiltonian is a Hamiltonian of
the form ${\cal H}=\sum_i H_i$, where each $H_i$ has bounded operator
norm: $\Vert H_i \Vert \leq J$ for some $J$, and where each $H_i$ has support
on a set of sites within some distance $R$ of site $i$.  This includes
lattice spin and fermion models with finite range interactions.

In general, we define a Hamiltonian to be local as follows:
we write the Hamiltonian as ${\cal H}=\sum_Z h_Z$
where $Z$ are different sets of sites on the lattice and $h_Z$ is
a hermitian operator with support on set $Z$.
For a Hamiltonian to be local, we require that there exist constants
$\mu,s_1$ such that for all $i$,
\be
\label{local}
\sum_{Z\ni i} \Vert h_Z \Vert |Z| \exp[\mu {\rm diam}(Z)] \leq s_1 < \infty,
\ee
where $|Z|$ is the total number of sites within set $Z$ (the cardinality).
For Goldstone's theorem we also impose an additional finite
dimensionality requirement, Eq.~(\ref{locl}), discussed below.

Eq.~(\ref{local}) implies a
Lieb-Robinson bound\cite{lr,hk,ns}: there exists
a velocity $v$ and constant $c$
such that for any two operators $A,B$ defined with
support on sets $X,Y$, 
\be
\label{lrb}
\Vert [A(t),B] \Vert \leq 
c \times |X| \Vert A \Vert \Vert B \Vert
\exp[-\mu {\rm dist}(X,Y)](\exp[v \mu |t|]-1),
\ee
where $A(t)=\exp[i {\cal H} t]
A \exp[-i {\cal H} t]$.  This bound is useful for $|t|\leq l/v$, and
implies that for any $\mu'\leq \mu$, we have
\be
\label{lrb2}
\Vert [A(t),B] \Vert \leq 
c \times |X| \Vert A \Vert \Vert B \Vert
\exp[-\mu'({\rm dist}(X,Y)],
\ee
so long as $t\leq l/v'$, with 
\be
\label{choice}
v'=v \mu/(\mu-\mu').
\ee
We will use the bound in the form (\ref{lrb2}) throughout.

Finally, we emphasize that our results are derived for finite volume systems.
The bounds that we find, however, for Goldstone's theorem, do not depend
on volume, while we will explicitly discuss the volume dependence for
the flux quantization result.  Stronger
Goldstone bounds may be possible in some cases in the infinite volume limit,
as there exist some systems for which, for example, there is a degenerate
ground state for a finite size system but a unique ground state in the
infinite system\cite{la2}; we do not consider these possibilities
further.  There also exist cases in which
there are two approximately degenerate low energy states, which
become exactly degenerate in the thermodynamic limit, with a gap
to the rest of the spectrum\cite{la1}; our results are applicable to
this case, as in the proof
of the Goldstone theorem we do not require exact degeneracy of
the low-lying states.

\section{Quasi-Adiabatic Continuation}
\subsection{Main Results}
The technique of quasi-adiabatic continuation\cite{zl,to} is a general
technique for describing how the ground state of gapped Hamiltonians evolves
under a change in a parameter.  We consider a parameter dependent
Hamiltonian ${\cal H}_s=\sum_Z h_Z(s)$ for some parameter $s$.  
We assume the ground
state or set of ground states is given at $s=0$ and we wish to describe the
evolution of this state or set of states under change in $s$.  We denote
the ground states of ${\cal H}_s$
by $\Psi_0^a(s)$ where $a=1...k$ is an index labeling
the $k$ distinct ground states.
The ground states $\Psi_0^a(s)$ are not required to be exactly degenerate:
they instead each have energy $E_a(s)$ and
we denote the splitting in energy between the highest and lowest energy
ground state by $\epsilon$.  If there is a degeneracy of the ground states
at some value of $s$, then there is some freedom in choosing the
ground state basis $\Psi_0^a$; we resolve this freedom by requiring that
if $E_a(s)=E_b(s)$ for some $s$,
then 
\be
\label{gaugechoice}
\langle \Psi_0^b(s),\partial_s \Psi_0^a(s) \rangle=0 \quad
{\rm whenever} \quad E_a(s)=E_b(s),
\ee
where $\langle ..., ...\rangle$ is used to denote the inner product.
Similarly, we require that
\be
\label{gaugechoice2}
\langle \Psi_0^a(s),\partial_s \Psi_0^a(s) \rangle=0
\ee
for all $a,s$.
We label other states of the system
by $\Psi^i$, $i>k$, with energy $E_i(s)$.

The most important
result of \cite{zl} shows that the evolution of the ground state
under a change in $s$ can be well approximated by a local Hermitian
operator acting on the ground state if there is a gap.
Let there be a gap $\Delta E$ between the sector of ground
states and the rest of the spectrum for all $s$.
Let $\Psi_0^a(s,s_Z)$ denote the set of ground states
of the Hamiltonian ${\cal H}(s)+(h_Z(s_Z)-h_Z(s))$.
Then it was shown that, for any length $l_Z$, there exists a Hermitian
operator ${\cal D}^{loc}_Z(s)$ and an anti-Hermitian matrix $N^{Z,loc}_{ba}$
such that 
$D^{loc}_Z(s)$ has support on the set of sites $j$ with
${\rm dist}(Z,j)\leq l_Z$ and satisfies
$\Vert {\cal D}^{loc}_Z(s) \Vert
\leq t_q \Vert \partial_{s_Z} h_Z(s_Z) \Vert$, and such that
\be
\label{qa1}
\Bigl|\partial_{s_Z} \Psi_0^a(s,s_Z)-
i{\cal D}^{loc}_Z(s) \Psi_0^a - \sum_b N^{Z,loc}_{ba}(s) \Psi_0^b(s)\Bigr| \leq
C_1
(\exp[-l_Z/\xi']+|Z| \exp[-\mu' l_Z])
t_q \Vert \partial_{s_Z} h_Z(s_Z) \Vert,
\ee
where $C_1$ is some numerical constant of order unity
(throughout, we denote such numerical constants by symbols $C_1,C_2,...$)
and we define
the constants $\xi',t_q$ by (the $q$ in $t_q$ stands for ``quasi-adiabatic"):
\begin{eqnarray}
\xi'=2v'/\Delta E,\\
t_q=\sqrt{l_Z/v'\Delta E}
\end{eqnarray}
In Eq.~(\ref{qa1}) and throughout this section,
all derivatives with respect to $s_Z$.
are taken at $s_Z=s$.

From this, we can define the operator ${\cal D}^{loc}(s)=\sum_Z {\cal D}^{loc}_Z(s)$
and bounds on $|\partial_s \Psi_0^a(s)-i{\cal D}^{loc}(s) \Psi_0^a -\sum_b
N^{loc}_{ba}(s) \Psi_0^b(s)|$ follow from Eq.~(\ref{qa1}), where
$N^{loc}_{ba}(s)=\sum_Z N^{Z,loc}_{ba}(s)$.

We briefly sketch the derivation of
Eq.~(\ref{qa1}) following ideas from \cite{zl}.  We define the operator
\be
\label{qa}
{\cal D}_Z(s)=i\int_0^{\infty} {\rm d}\tau e^{-(\tau/t_q)^2/2}
[\tilde w_{Z,s_Z}^+(i\tau)-h.c.],
\ee
where $w_{Z,s_Z}=(\partial_{s_Z} h_Z(s_Z))$ and
where for any operator $A$ we define, following \cite{loc}:
\begin{eqnarray}
\label{tld}
\tilde A(t)\equiv A(t) \exp[-(t/t_q)^2/2], \\
\tilde A^{\pm}(i\tau)=\frac{1}{2\pi}\int {\rm d}t \tilde A(t) \frac{1}{\pm i t+\tau}.
\end{eqnarray}
The time evolution of operators in Eq.~(\ref{tld}) is the usual Heisenberg
evolution with the Hamiltonian ${\cal H}(s)$.
Define
\be
\label{nba}
N_{ba}^Z(s)=\langle \Psi_0^b(s),\partial_{s_Z} \Psi_0^a(s,s_Z) \rangle
-\langle \Psi_0^b(s),i{\cal D}_Z(s) \Psi_0^a(s) \rangle.
\ee
Then, one may show using the gap that for any $\Psi_0^a$ we have
\begin{eqnarray}
\label{b0}
&&\Bigl|\partial_{s_Z}\Psi_0^{a}(s,s_Z)
-i{\cal D}_Z(s) \Psi_0^a(s)
-\sum_b N_{ba}^Z(s) \Psi_0^b\Bigr|\\ \nonumber &=&
\Bigl|-\sum_{i\neq a} \frac{1}{E_i(s)-E_a(s)}
\Bigl( \langle \Psi^i,\partial_{s_Z} h_Z(s_Z) \Psi_0^a \rangle \Bigr)
\Psi^i-i{\cal D}_Z(s)\Psi_0^a(s,s_Z)-\sum_b N_{ba}^Z(s) \Psi_0^b(s)\Bigr|\\ \nonumber
&\leq&
C_2(t_q \exp[-(t_q \Delta E)^2/2])\Vert \partial_{s_Z}
h_Z(s_Z) \Vert).
\end{eqnarray}

We define ${\cal D}_Z^{loc}(s)$ as follows: define $U_l$ to be a unitary
operator on the set of sites $j$ such that ${\rm dist}(Z,j)>l$.  Let
$\mu(U_l)$ be the Haar measure on such operators.
Define\cite{lightcone}
\be
{\cal D}^{loc}_Z(s)=\int {\rm d}\mu(U_{l_Z}) U_{l_Z}^{\dagger} {\cal D}_Z(s) U_{l_Z}.
\ee
Using Lieb-Robinson bounds, it is possible to show that
\be
\label{db}
\Vert {\cal D}^{loc}_Z(s)-{\cal D}_Z(s) \Vert \leq
C_3(\exp[-(l_Z/v't_q)^2/2]+|Z| \exp[-\mu' l_Z]) t_q \Vert
\partial_{s_Z} h_Z(s_Z) \Vert.
\ee
Combining these results (\ref{b0},\ref{db})
gives Eq.~(\ref{qa1}), where $N_{ba}^{Z,loc}(s)=
\langle \Psi_0^b(s), \partial_{s_Z} \Psi_0^a(s,s_Z) \rangle-
\langle \Psi_0^b(s),i{\cal D}_{Z,loc}(s) \Psi_0^a(s) \rangle$.
Define the operator $N^{Z,loc}(s)=\sum_{ba} 
N_{ba}^{Z,loc}(s)
\Psi_0^b\rangle\langle\Psi_0^a$ and
$N^{loc}(s)=\sum_{ba} 
N_{ba}^{loc}(s) \Psi_0^b\rangle\langle\Psi_0^a$.

\subsection{Berry Phase}
The discussion that follows in this subsection is meant to introduce
the idea of computing a Berry phase for quasi-adiabatic evolution
and show that it is close to the usual definition of the Berry phase.
This discussion is not used in the rest of the paper and may be
skipped if desired, but seems to be interesting in itself as well as for
other future applications.

Define
$P_0(s)$ to be the projector onto the ground state
sector: $P_0(s)\equiv\sum_{a=1}^k \Psi_0^a(s)\rangle
\langle \Psi_0^a(s)$.
A direct computation shows that for all $a,b$
\be
\label{n}
|\langle \Psi_0^b(s), {\cal D}_{Z,loc}(s) \Psi_0^a(s)\rangle| \leq 
C_4 (t_q \exp[-l_Z/\xi']+t_q |Z| \exp[-\mu' l_Z]+t_q^2 \epsilon) s
\Vert \partial_{s_Z}h(s_Z) \Vert.
\ee
Note that for large $l_Z$ and small $\epsilon$, the right-hand side
of this equation becomes close to zero, giving a result similar to
Eqs.~(\ref{gaugechoice},\ref{gaugechoice2}); this will be used to
relate the evolution in the ground state sector produced by
$i{\cal D}(s)$ to a Berry phase.

Eq.~(\ref{qa1}) gives the evolution of wavefunctions
under an infinitesimal change in $s$.  We now
present a unitary operator $V$ to describe the change in wavefunctions
from $s=0$ to $s=1$.  We define
\be
\label{expqad}
V(s)\equiv {\cal S}' \exp[\int_0^s {\rm d}s' i {\cal D}(s')],
\ee
where ${\cal S}'$ denotes that the exponential is $s'$-ordered.
Then, $V\Psi_0^a(0)$ is close in norm to some linear combination
of states $\Psi_0^b(s)$, with an error that may be estimated from
Eq.~(\ref{qa1}).
If the Hamiltonian is quasi-adiabatically continued around a closed
path in parameter space, so that ${\cal H}(s_{final})={\cal H}(0)$,
then $|V(s_{final})\Psi_0^a-
\sum_{b} Q_{ba}\Psi_0^b|\leq
C_1\sum_Z(\exp[-l_Z/\xi']+|Z| \exp[-\mu' l_Z])t_q
s_{final} \Vert \partial_{s_Z}h(s_Z) \Vert$
where
\begin{eqnarray}
Q&=&{\cal S}' \exp[\int_0^s {\rm d}s' i P_0(s'){\cal D}(s')P_0(s')-
P_0 \sum_a (\partial_{s'} \Psi_0^a(s'))\rangle\langle \Psi_0^a(s')]
\\ \nonumber
&=&{\cal S}' \exp[-\int_0^s {\rm d}s' N^{loc}(s')].
\end{eqnarray}
Eq.~(\ref{n}) gives
\be
\Vert Q-Q^{berry}\Vert \leq
C_4\sum_Z(t_q \exp[-l_Z/\xi']+t_q |Z| \exp[-\mu' l_Z]+t_q^2 \epsilon)
s_{final} \Vert \partial_{s_Z}h(s_Z) \Vert,
\ee
where $Q^{berry}$ is the non-Abelian Berry phase for an adiabatic
evolution in a related system in which the level splitting between
the ground states $\Psi_0^a(s)$ is set equal to zero\cite{berry}.

While we used the assumption that there is a gap $\Delta E$ to derive
Eq.~(\ref{b0}), it
is important to note that this equation in fact depends only on the
weaker
assumption that the state $\partial_s h_Z(s)\Psi_0^a$ is a linear combination
of ground states and states $\Psi^i$ with $E_i\geq \Delta E+E_a$ for $a=1...k$.
This follows automatically if there is a gap in the spectrum, but it also
follows in a spin system 
if, for example, there is a gap $\Delta E$ between the ground
state with spin-$0$ and the lowest spin-$1$ state, and $\partial_s h_Z(s)$
has spin-$1$.

\section{Goldstone's Theorem in Spin-Gapped Systems}
\subsection{Statement of Result and Examples}
Recently, the exponential {\it clustering} of correlation functions
in a system with a gap has been proven\cite{lsmhd,loc,hk,ns}.  That is,
it has been shown that, for a local Hamiltonian with a gap $\Delta E$
between a sector of degenerate ground states, labeled $\Psi_0^a$ for
$a=1...k$,
and the rest of the spectrum,
the connected correlated function,
$\langle A B \rangle - \langle A P_0 B \rangle$,
of two operators $A,B$ with support on sets $X,Y$
is exponentially small in the distance ${\rm dist}(X,Y)$.  Here,
$\langle O \rangle \equiv k^{-1} \sum_{a=1}^k
\langle \Psi_0^a, O \Psi_0^a \rangle$ and
$P_0\equiv\sum_{a=1}^k \Psi_0^a \rangle\langle \Psi_0^a$ is
the projector on the ground state sector.
For a parameter dependent
Hamiltonian, we define
$\langle O \rangle_s \equiv k^{-1} \sum_{a=1}^k
\langle \Psi_0^a(s), O \Psi_0^a(s) \rangle$.  In the event
that the states $\Psi_0^a$
are not exactly degenerate but only
approximately degenerate, similar bounds were found with corrections
that depend on the energy difference between the states.

In this section we prove a stronger
statement about the decay of
correlation functions in gapped systems with a continuous symmetry.
We consider Hamiltonians which obey Eq.~(\ref{local}) for some $\mu,s_1$
and we also require that
for any site $i$ the number of sites $j$ within distance $l$ of site
$i$ is bounded by $a l^d$ for some finite $a,d$:
\be
\label{locl}
\sum_{j,{\rm dist}(i,j)\leq l} 1 \leq a l^d.
\ee
We consider Hamiltonians which have a $U(1)$ symmetry as follows:
for each site $i$ we assume that there exists a local operator $q_i$
with support on that site such that $[Q,{\cal H}]=0$ where $Q=\sum_i q_i$.
For any set $X$, we define $R(\theta,X)=\prod_{i\in X} \exp[i q_i \theta]$.
We consider operators $\phi_X,\overline \phi_Y$ with support on sets $X,Y$
which transform as vectors as follows under this $U(1)$ symmetry:
$R(-\theta,X) \phi_X R(\theta,X)=\exp[i \theta] \phi_X$ and
$R(-\theta,Y) \overline \phi_Y R(\theta,Y)=\exp[-i \theta] \overline \phi_Y$.
We assume that $\Vert q_i \Vert \leq q_{max}$ for some $q_{max}$.
We will show that, for a Hamiltonian which obeys all these
conditions and has a gap $\Delta E$
between the ground state
sector and the rest of the spectrum
\begin{eqnarray}
\label{claim}
\langle \phi_X \overline \phi_Y \rangle
&\leq &
C_5 \Vert \phi_X \overline \phi_Y \Vert
|X| a {\rm dist}(X,Y)^d s_1 q_{max}
\sqrt{{\rm dist}(X,Y)\Delta E/v'}
\times \\ \nonumber &&(\exp[-\mu'{\rm dist}(X,Y)/2]+\exp[-{\rm dist}(X,Y)/2\xi']).
\end{eqnarray}
Picking the optimal value of $\mu'$ in Eq.~(\ref{choice}), we show that
\begin{eqnarray}
\label{claim2}
\langle \phi_X \overline \phi_Y \rangle
\leq 
C_6 \Vert \phi_X \overline \phi_Y \Vert
|X| {\rm dist}(X,Y)^d q_{max} s_1
\sqrt{{\rm dist}(X,Y)\Delta E/v'}
\exp[-{\rm dist}(X,Y)/2\xi],
\end{eqnarray}
with
\be
\xi=2v/\Delta E+\mu.
\ee

We emphasize that
we do not require the states in the ground state sector to be degenerate with
each other, simply the existence of a gap between that sector and the rest
of the spectrum.  Further, the bounds do not depend on the
value of $k$ or on energy difference between the states $\Psi_0^a$.
This result (\ref{claim},\ref{claim2})
is stronger than that in \cite{k2,hk} which
was only valid for $d<2$ while the present result
is valid in arbitrary dimension.  It is also stronger than
other previous results\cite{ergodic} which either required a unique
ground state or else assumed an ergodic property which is equivalent
to requiring the vanishing of the matrix elements in the ground state
sector; in either of these cases the decay of correlations becomes equivalent to
clustering.

Before giving the proof we motivate the definition of $q_i$ by discussing
physical examples.
In a Bose system with conserved particle number, the $q_i$ can represent
the particle number on a given site and the operators $\phi_X,\overline \phi_Y$
can represent creation and annihilation operators for the bosons.  For a
spin system, the $q_i$ can represent the $z$ component of the spin on a site
and the $\phi_X,\overline \phi_Y$ can represent raising and lowering spin
operators on sites.  Consider such
a spin system in a disordered quantum paramagnetic
phase.  An example of such a system would be obtained by a two
dimensional Hamiltonian of spin-$1/2$ spins
for a system of two layers with coordinates
labeled $(i,l)$ where $l=1,2$ is a layer label and $i$ indexes position
in each layer.  First consider the Hamiltonian:
\be
{\cal H}=J\sum_{i,j\, n.n.} \vec S_{i,1} \cdot \vec S_{j,1}
+J_{\perp} \sum_i \vec S_{i,1} \cdot \vec S_{i,2},
\ee
with $J,J_{\perp}>0$ where the first sum is over nearest neighbor $i,j$.
This is a bilayer model Heisenberg model that is connected with
the Kondo lattice model\cite{blyr}.
For $J_{\perp}>>J$, the ground state is unique and gapped, with spins
$(i,1),(i,2)$ being in a singlet state with high probability.  In this
case, Eq.~(\ref{claim}) shows  $\langle S^x_{i,1} S^x_{j,1} \rangle$
is exponentially decaying in ${\rm dist}(i,j)$ but this provides
no additional information beyond that already known from the exponential
decay of correlation functions.  Now consider the Hamiltonian on a lattice
with two ``defects": for some given $k,l$, we replace the spins at sites
$(k,2)$ and $(l,2)$ with spin-$0$ spins\cite{blyrdefect}.  
That is, we remove them from
the lattice, so $\vec S_{k,2}=\vec S_{l,2}=0$.  Then, there are unpaired
spins at $(k,1)$ and $(l,1)$.  
In the limit $J=0$, indeed there are four exactly
degenerate ground states.
Returning to the case $J_{\perp}>>J>0$, we
expect that if ${\rm dist}(k,l)$
is large
then there will be four low energy states and then a gap to the
rest of the spectrum.  In this case, Eq.~(\ref{claim}) can be used, given
the assumption of a gap as all other requirements are trivially satisfied,
to bound $\langle S^x_{k,1} S^x_{l,1} \rangle$ and hence while the operators
$\vec S_{k,1},\vec S_{l,1}$
may have nonvanishing matrix elements between these states, the {\it average}
of the correlation function over the different ground states is small

\subsection{Proof}
To show Eq.~(\ref{claim}), we define a set of parameter dependent Hamiltonians 
${\cal H}_{\theta}=\sum_Z h_Z(\theta)$ as follows.
Let $X'$ denote the set of sites $i$
such that ${\rm dist}(X,i)\leq {\rm dist}(X,Y)/2$, as shown
in the figure.  Then define
$h_Z(\theta)=R(X',\theta) h_Z R(X',-\theta)$.
Clearly, as $R(X',-\theta)$ is a unitary transformation,
${\cal H}_{\theta}$ has the same spectrum of ${\cal H}$ and
the ground states of ${\cal H}_{\theta}$ are given by
$\Psi_0^a(\theta)=R(X',\theta) \Psi_0(\theta)$.

\begin{figure}[tb]
\centerline{
\includegraphics[scale=0.7]{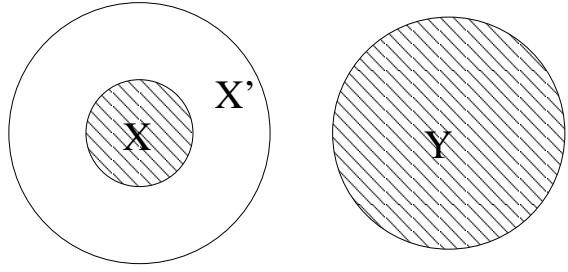}}
\caption{Illustration of the geometry we consider.  $X,Y$ are shown as shaded regions, while $X'$ includes everything within the outer circle around $X$.}
\end{figure}

Thus, 
\begin{eqnarray}
\label{cwt}
\partial_{\theta} \langle \phi_X \overline \phi_Y \rangle_{\theta}
&=&\partial_{\theta} \langle R(X',-\theta) \phi_X \overline \phi_Y 
R(X',\theta) \rangle 
\\ \nonumber
&=&\partial_{\theta} \exp[i \theta] \langle \phi_X \overline \phi_Y 
\rangle 
=i \langle \phi_X \overline \phi_Y 
\rangle,
\end{eqnarray}
where we used the fact that $X\subset X'$ while $Y\cap X'=0$ so that
$[\overline \phi_Y,R(X',\theta)]=0$ and where we evaluate the derivatives
at $\theta=0$.

Without loss of generality we may assume
that 
\be
\label{localcon}
[Q,h_Z]=0
\ee
for all $Z$.  To show this, we use $[Q,{\cal H}]=0$ to
write
${\cal H}=\sum_Z O_Z$ where $O_Z\equiv (1/2\pi)
\int_0^{2\pi} {\rm d}\theta \, \exp[i Q \theta]
h_Z \exp[-i Q \theta]$.  Then, $O_Z$ has support on $Z$ and $\Vert O_Z \Vert
\leq \Vert h_Z \Vert$.  Finally, $[Q,O_Z]=0$.  Hence, by replacing $h_Z$
by the operators $O_Z$, we succeed in rewriting the Hamiltonian such that
Eq.~(\ref{localcon}) is satisfied and such that the Hamiltonian still obeys
Eq.~(\ref{local}).
From Eq.~(\ref{localcon}),
\begin{eqnarray}
\label{onborder}
Z\cap X'=0&\rightarrow&h_Z(\theta)=h_Z,\\ \nonumber
Z\subseteq X'&\rightarrow&h_Z(\theta)=h_Z .
\end{eqnarray}
That is, $h_Z(\theta)=h_Z$ unless $Z$ contains some point $i \in X'$ and
some other point $j \not \in X'$.

We now use quasi-adiabatic continuation to bound
the first line of Eq.~(\ref{cwt}), thus bounding the correlation
function on the last line of this equation.
Using 
$\partial_{\theta} \langle \phi_X \overline \phi_Y \rangle_{\theta}
=
\frac{1}{k}\sum_{a=1}^k
\partial_{\theta}
\langle \Psi_0^a(\theta), \phi_X \overline \phi_Y \Psi_0^a(\theta) \rangle$
and 
Eq.~(\ref{qa1}) we have
\begin{eqnarray}
\label{cwd}
\Bigl|\partial_{\theta} \langle \phi_X \overline \phi_Y \rangle_{\theta}\Bigr|
&\leq&
\Bigl| \sum_Z \frac{1}{k}\sum_{a=1}^k
\Bigl(
i \langle \Psi_0^a(\theta), [\phi_X \overline \phi_Y ,{\cal D}^{loc}_Z(\theta=0)]
\Psi_0^a(\theta) \rangle+\\ \nonumber
&&\sum_{b=1}^k
N_{ba}^{Z,loc}(\theta=0)
\langle \Psi_0^a(\theta), \phi_X \overline \phi_Y
\Psi_0^b(\theta) \rangle
+\sum_{b=1}^k N_{ba}^{Z,loc}(\theta=0)^*
\langle \Psi_0^b(\theta), \phi_X \overline \phi_Y
\Psi_0^a(\theta) \rangle\Bigr)\Bigr|\\ \nonumber
&&+ C_1
\sum_Z (\exp[-l_Z/\xi']+|Z| \exp[-\mu' l_Z])
t_q \Vert \partial_{\theta_Z} h_Z(\theta_Z) \Vert
\Vert \phi_X \overline \phi_Y \Vert,
\end{eqnarray}
The terms involving the matrix $N^{Z,loc}$ in Eq.~(\ref{cwd}) cancel after
summing over $a$ and $b$.
This is the key step which makes the quasi-adiabatic continuation
useful in deriving the desired exponential decay of correlations.
We will now show that all the other terms on the right-hand
side of Eq.~(\ref{cwd}) are exponentially
small in ${\rm dist}(X,Y)$, when we choose
$l_Z={\rm min}({\rm dist}(X,Z),{\rm dist}(Y,Z))-1$.  This requires
a straightforward, but lengthy, series of inequalities.

The commutator
$[\phi_X \overline \phi_Y ,{\cal D}^{loc}_Z(\theta=0)]$ is vanishing unless
${\cal D}^{loc}_Z(\theta=0)$ has support on either set $X$ or set $Y$ which,
with the above choice of $l_Z$, occurs only if $Z\cap X\neq \emptyset$
or $Z\cap Y\neq \emptyset$.
Further, from Eq.~(\ref{onborder}) ${\cal D}^{loc}_Z(\theta=0)$ is vanishing
unless $Z$ includes some point $i\in X'$ and some other point $j\not\in X'$.
Thus, the commutator
$[\phi_X \overline \phi_Y ,{\cal D}^{loc}_Z(\theta=0)]$ is vanishing unless either
$Z$ includes some point $i\in X$ and some other point
$j \not \in X'$ or else $Z$ includes
some point $i\in X'$ and some other point
$j\in Y$.  In either case, we have $i \in X'$ and
${\rm dist}(i,j)\geq {\rm dist}(X,Y)/2$.
Thus, 
\begin{eqnarray}
\label{comwD}
\sum_Z
\Vert [\phi_X \overline \phi_Y ,{\cal D}^{loc}_Z(\theta=0)] \Vert &\leq&
2 t_q \Vert \phi_X \overline \phi_Y \Vert \sum_{i \in X'} \quad
\sum_{j, {\rm dist}(i,j)\geq {\rm dist}(X,Y)/2} \quad
\sum_{Z\ni i,j}
\Vert \partial_{\theta_Z} h_Z(\theta_Z) \Vert 
\\ \nonumber
&\leq &
4 t_q q_{max} \Vert \phi_X \overline \phi_Y \Vert 
|X'| s_1 \exp[-\mu' ({\rm dist}(X,Y)/2)]\\ \nonumber
&\leq&
4 t_q q_{max} \Vert \phi_X \overline \phi_Y \Vert 
a ({\rm dist}(X,Y)/2)^d
|X| s_1 \exp[-\mu' ({\rm dist}(X,Y)/2)],
\end{eqnarray}
where we used Eqs.~(\ref{local},\ref{locl}) and
$\Vert {\cal D}_Z^{loc} \Vert \leq t_q \Vert \partial_{\theta_Z} h_Z(\theta_Z)
\Vert$ and
$\Vert \partial_{\theta_Z} h_Z(\theta_Z) \Vert \leq
2 q_{max} \Vert h_Z \Vert$ for the last two inequalities.

Similarly, noting that $l_Z \geq {\rm dist}(X,Y)/2-{\rm diam}(Z)$,
\begin{eqnarray}
\label{errterms}
&&\sum_Z
\Bigl|(\exp[-l_Z/\xi']+|Z| \exp[-\mu' l_Z])\Bigr|
t_q \Vert \partial_{\theta_Z} h_Z(\theta_Z) \Vert
\\ \nonumber
&\leq&
\sum_{i \in X'} \sum_{Z \ni i}
2 \Bigl(\exp[-({\rm dist}(X,Y)/2-{\rm diam}(Z))/\xi']+|Z| \exp[-\mu' 
({\rm dist}(X,Y)/2-{\rm diam}(Z)]\Bigr) t_q q_{max} \Vert h_Z \Vert
\\ \nonumber
&\leq& |X| a({\rm dist}(X,Y)/2)^d s_1 q_{max}
\sqrt{{\rm dist}(X,Y)\Delta E/v'}
\Bigl(\exp[-\mu'{\rm dist}(X,Y)/2]+\exp[-{\rm dist}(X,Y)/2\xi']\Bigr).
\end{eqnarray}

Combining Eqs.(\ref{comwD},\ref{errterms}), we find that
\begin{eqnarray}
|\partial_{\theta} \langle \phi_X \overline \phi_Y \rangle_{\theta}|
&\leq&
C_5 \Vert \phi_X \overline \phi_Y \Vert
|X| a {\rm dist}(X,Y)^d s_1 q_{max}
\sqrt{{\rm dist}(X,Y)\Delta E/v'}
\times\\ \nonumber &&(\exp[-\mu'{\rm dist}(X,Y)/2]+\exp[-{\rm dist}(X,Y)/2\xi']).
\end{eqnarray}
Combining this with Eq.~(\ref{cwt}), we arrive at Eq.(\ref{claim}), as desired.

It is interesting to note that Eq.~(\ref{locl}) is necessary in this
derivation.
We sketch a system for which Eq.~(\ref{locl}) does not hold, and
show how a Goldstone theorem may fail in this case.
Consider a random graph with $V$ nodes each having coordination number
$3$.
Consider a set of $V$ spin-$1/2$ spins, with Hamiltonian
\be
H=-\sum_{i,j} J_{ij} \vec S_i \cdot \vec S_j,
\ee
where the interaction matrix $J_{ij}$ equals $1$ if the nodes $i,j$
are connected by an edge on the graph, and zero otherwise.  The interaction
is ferromagnetic, so pointing all spins up (or in any other direction)
gives a ground state.  Further, the Hamiltonian is local, using a shortest
path metric on the graph to define ${\rm dist}(i,j)$.
However, a random graph of this form is typically an expander 
graph\cite{expander} with a gap in the spectrum of the
graph Laplacian,
so a spin-wave theory calculation\cite{spinwave}
gives a gap in the magnon spectrum.  Thus, this system has a set of degenerate
ground states and a gap.  However, the spin correlations do not decay,
as $\langle \vec S_i \cdot \vec S_j \rangle=1/4$ for all $i\neq j$.

\section{Flux Quantization in Spin-Gapped Systems}
\subsection{Examples and Statement of Result}
In this section we consider flux periodicity in systems of fermions with
half-integer spin.
Before stating the precise result we motivate some of the definitions
through physical examples.  Consider a system of fermions on a cylinder,
and suppose that an additional magnetic field is applied through the cylinder,
introducing an Aharonov-Bohm phase for the fermions.
The Hamiltonian is unchanged, up to a gauge transformation,
if the magnetic flux through the cylinder
is changed by $2\pi$, and hence the ground state energy is
periodic in the flux with periodic $2\pi$.  For an arbitrary Hamiltonian
the energy of the ground state
for $\pi$ flux may be much different for that at $0$ flux.
However, a superconductor typically has its ground state
energy approximately periodic in the flux with period $\pi$ due to
so-called Byer-Yang states\cite{by}, up to some
corrections which are exponentially small in system size $L$.
This is often taken to imply that the elementary excitation in the system
has charge $2e$ rather than $e$.

However, there are other interesting scenarios in which the
ground state energy at $\pi$ flux is close to that at zero flux.
Consider a system at half filling which forms a resonating
valence bond state (RVB) with a spin gap.  Then dope this system by introducing
a small number of holes.  The holes have vanishing spin and charge $+e$, and
may Bose condense, again giving a superconductor.  Naively, one may
expect that in this case the ground state energy of the system at
$\pi$ flux would be very different from that at zero flux as the excitations
that condense have charge $e$.
However, the presence of topological excitations of the valence bond
system implies that in fact the ground state energy is still approximately
periodic in the flux with period $\pi$\cite{e2e,others,luttinger}.  The
question was raised in \cite{e2e} whether there were some general conditions
under which one could prove that this approximate periodicity held for
fermionic system.  In this section, we provide a partial answer
to this question by considering the case in which
the system has a {\it spin gap}, as would occur for an $s$-wave superconductor
or for a short wave RVB state, and prove that there exists a state of
the system at $\pi$ flux which is close to the ground state
energy at zero flux.  Unfortunately, our results will not apply to
$d$-wave superconductors where there are gapless spin excitations.

We now state the result.
We assume that there are conserved spin and charge as follows.
On each site $i$ we assume that there is a Hilbert with
dimension 4, representing the states of no fermions, one fermion with spin
up, one fermion with spin down, and two fermions.
We define the 
spin operator on site $i$, $S^z_i$, to be diagonal in this basis with
eigenvalues $0,+1/2,-1/2,0$ respectively, and we
define the fermion number operator on site $i$, $q_i$, to be
diagonal in this basis with eigenvalues $0,1,1,2$.
We assume that the Hamiltonian
${\cal H}$ commutes with the operators $S^z=\sum_i S^z_i$ and
$Q=\sum_i q_i$; as before, this means that we may assume without
loss of generality that
$[S^z,h_Z]=[Q,h_Z]=0$ for all $Z$.
Note that $S^z_i\pm q_i/2$ has integer eigenvalues.
We assume that the Hamiltonian is local in the sense of Eq.~(\ref{local})
and we assume that the system is periodic in one direction
as follows: let the system be defined on a $d$ dimensional lattice of $V$
different sites, with any
site $i$ having coordinates
$(x_0(i),x_1(i),...x_{d-1}(i))$, with periodic boundary conditions in the
$x_0$ direction with period $L$.  Assume that ${\rm dist}(i,j)\geq
\min_n |x_0(i)-x_0(j)-nL|$, where the minimum is over all integers
$n$.  This is satisfied by any physically reasonable metric on the lattice,
such as a Manhattan metric.

We assume that ${\cal H}$ has a unique ground state $\Psi_0$, 
although the results can
be readily extended to a system with multiple ground states with small
splitting $\epsilon$.
We define a Hamiltonian ${\cal H}$ 
to have a spin gap $\Delta E$ if the the ground state has vanishing spin and
the first excited state with non-vanishing spin has energy at least $\Delta E$
above the ground state energy.

We now define the Hamiltonians ${\cal H}_{\theta}$ with flux $\theta$ inserted.
Define ${\cal L}$ to be the set of all points $i$ with $0\leq x_0(i)<L/2$.
For any set $X$, define $R_q(\theta,X)=\prod_{i \in X} \exp[i q_i \theta]$.
Define
${\cal H}_{\theta}=
\sum_Z
h_Z(\theta)$,
where we define $h_Z$ as follows:
if $Z$ contains any
points $i$ with $L/4\leq x_0(i) \leq 3L/4$, then we define
\be
h_Z(\theta)=
R_q(-\theta)
h_Z
R_q(\theta)
\ee
Otherwise, we define
\be
h_Z(\theta)=h_Z.
\ee

We will prove that, under the conditions identified above,
(conserved spin and charge,
half-integer spin, locality, periodicity, uniqueness of ground state,
and spin gap), that there exists some state, which we will write
as $W_1(\pi)\Psi_0$ with $W_1$ a unitary operator, such that
the difference in energies
\begin{eqnarray}
\label{qe}
|\langle W_1(\pi) \Psi_0, {\cal H}_{\pi}
W_1(\pi)\Psi_0\rangle-\langle \Psi_0, {\cal H} \Psi_0 \rangle|
\leq C_7 V^2 \sqrt{L/v \Delta E} s_1^2
\exp[-L/4(\xi+1/\mu)].
\end{eqnarray}
Note that $W_1(\pi) \Psi_0$ need not be the ground state of ${\cal H}_{\pi}$.

\subsection{Proof}
In this subsection we provide the proof.  The main idea is to define
separate fluxes for up and down fermions separate, and use quasi-adiabatic
continuation to carry one flux from $0$ to $\pi$ and the other from $0$ to
$-\pi$.  Then, the proof closely follows the proof of a higher dimensional
Lieb-Schultz-Mattis system given in\cite{suffcon}.
For any set $X$, define $R_{\uparrow}(\theta,X)=\prod_{i\in X}
\exp[i (S^z_i+q_i/2) \theta]$ and
$R_{\downarrow}(\theta,X)=\prod_{i\in X}
\exp[i (-S^z_i+q_i/2) \theta]$.  
Thus, $R_{\uparrow}$ produces a gauge transformation on the
up spin fermions and $R_{\downarrow}$ transforms the down spin fermions.
Note that $R_{\uparrow}(\pi,X) R_{\downarrow}(-\pi,X)=
R_q(\pi,X)$; this equality depends on the fact that
$R_{\downarrow}(-\pi,X)=R_{\downarrow}(\pi,X)$.

\begin{figure}[tb]
\centerline{
\includegraphics[scale=0.7]{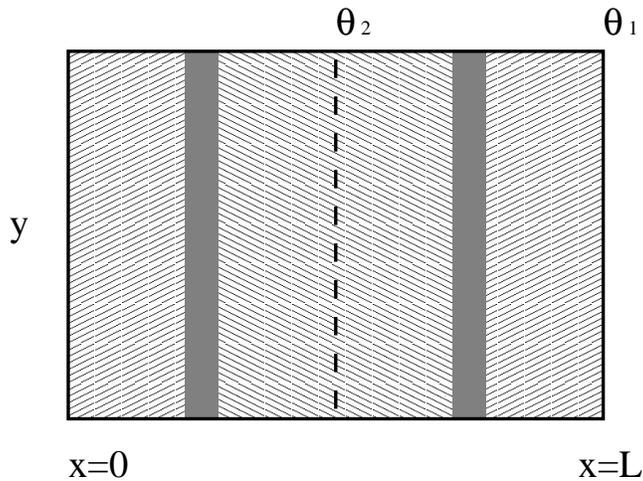}}
\caption{Illustration of the geometry we consider.  Flux
is inserted along the dashed line in the center, and at the edge.
The slanting upwards line is the support of $W_1$, the slanting downwards
is the support of $W_2$.  The grey areas have width $L/2(\mu\xi+1)$.  The
lines $x_0(i)=L/4$ and $x_0(i)=3L/4$ pass vertically through the middle
of the left and right grey areas, respectively.}
\end{figure}

We now define a family of Hamiltonian with separate up and down fluxes
inserted, and we also insert flux at two different points: along a line with
$x_0=0$ and along a line with $x_0=L/2$ as shown in the figure.
By gauge invariance the spectrum
of the Hamiltonian depends only on the total of these fluxes, but the
introduction of two different flux angles is a
useful technical device.  We define four flux angles, $\theta_1^{\uparrow},
\theta_1^{\downarrow}, \theta_2^{\uparrow}, \theta_2^{\downarrow}$, where
the fluxes described by angles
$\theta_1$ are along the line with $x_0=0$ and the fluxes described by
angles $\theta_2$ are
along the line with $x_0=L/2$, and where we introduce different fluxes
for up and down particles.  The insertion of physical magnetic flux into
into the system the system leads to $\theta_1^{\downarrow}=\theta_1^{\uparrow}$
and $\theta_2^{\downarrow}=\theta_2^{\uparrow}$, however our introduction
of separate angles for up and down particles is necessary to derive the
final result.
The resulting Hamiltonian 
${\cal H}(\theta_1^{\uparrow},\theta_1^{\downarrow},
\theta_2^{\uparrow},\theta_2^{\downarrow})=\sum_Z
h_Z(\theta_1^{\uparrow},\theta_1^{\downarrow},
\theta_2^{\uparrow},\theta_2^{\downarrow})$, where we define $h_Z$ as follows:
if $Z$ contains any
points $i$ with $L/4\leq x_0(i) \leq 3L/4$, then we define
\be
h_Z(\theta_1^{\uparrow},\theta_1^{\downarrow},
\theta_2^{\uparrow},\theta_2^{\downarrow})=
R_{\uparrow}(-\theta_2^{\uparrow},{\cal L})
R_{\downarrow}(-\theta_2^{\downarrow},{\cal L})
h_Z
R_{\uparrow}(\theta_2^{\uparrow},{\cal L})
R_{\downarrow}(\theta_2^{\downarrow},{\cal L}).
\ee
Otherwise, we define
\be
h_Z(\theta_1^{\uparrow},\theta_1^{\downarrow},
\theta_2^{\uparrow},\theta_2^{\downarrow})=
R_{\uparrow}(\theta_1^{\uparrow},{\cal L})
R_{\downarrow}(\theta_1^{\downarrow},{\cal L})
h_Z
R_{\uparrow}(-\theta_1^{\uparrow},{\cal L})
R_{\downarrow}(-\theta_1^{\downarrow},{\cal L}).
\ee
Then, if $\theta_1^{\uparrow}=-\theta_2^{\uparrow}$ and
$\theta_1^{\downarrow}=-\theta_2^{\downarrow}$, we have
$\Psi_0(\theta_1^{\uparrow},\theta_1^{\downarrow},
-\theta_1^{\uparrow},-\theta_1^{\downarrow})=
R_{\uparrow}(\theta_1^{\uparrow},{\cal L})
R_{\downarrow}(\theta_1^{\downarrow},{\cal L})
\Psi_0$.  Further, the eigenvalues of the Hamiltonian 
${\cal H}(\theta_1^{\uparrow},\theta_1^{\downarrow},
\theta_2^{\uparrow},\theta_2^{\downarrow})$
depend only
on $\theta_1^{\uparrow}+\theta_2^{\uparrow}$ and
$\theta_1^{\downarrow}+\theta_2^{\downarrow}$.
We define ${\cal Z}_c$ to be the set of $Z$ such that 
$h_Z(\theta_1^{\uparrow},\theta_1^{\downarrow},
\theta_2^{\uparrow},\theta_2^{\downarrow})$ is different from $h_Z$.
Crucially, the physical Hamiltonian with $\pi$ magnetic flux inserted is
\be
\label{piispi}
{\cal H}_{\pi}={\cal H}(\pi,\pi,0,0).
\ee

The proof now closely follows\cite{suffcon}, and we only sketch the
proof; the only differences from \cite{suffcon} are the presence of
separate flux angles for up and down fermions, and some minor
complications due to exponentially decaying interactions, rather than
finite range.  We define $W_1(\phi)$ to be the operator which
quasi-adiabatically continues ${\cal H}(\theta,-\theta,0,0)$ from
$\theta=0$ to $\theta=\phi$, using Eq.~(\ref{expqad}) with $l_Z=L/4-{\rm diam}(Z)-L/4(\xi\mu+1)$.  We define
$W_2(\phi)$ to be the operator which quasi-adiabatically
continues ${\cal H}(0,0,\theta,-\theta)$ from $\theta=0$ to
$\theta=-\phi$.  We define $W(\phi)$ to be the
operator which quasi-adiabatically continues
${\cal H}(\theta,-\theta,-\theta,\theta)$ from $\theta=0$ to
$\theta=\phi$.

Following the results in \cite{suffcon} one may show that
\begin{eqnarray}
\label{close}
|R_{\uparrow}(\pi,X) R_{\downarrow}(-\pi,X) \Psi_0-W(\pi) \Psi_0|
\leq c_2,
\end{eqnarray}
where 
\begin{eqnarray}
c_2&=&C_8\sum_{Z, Z\in {\cal Z}_c} t_q \Vert h_Z \Vert
(\exp[-l_Z/\Delta E/2v']+|Z| \exp[-\mu' l_Z])
\\ \nonumber
&\leq&
C_8\sqrt{L/v\Delta E} s_1 V
(\exp[-l_Z/\Delta E/2v']+|Z| \exp[-\mu' l_Z])
.
\end{eqnarray}
and also that
\begin{eqnarray}
\label{commute}
\Vert W_1(\pi) W_2(\pi)-W(\pi)\Vert 
\leq c_3,
\end{eqnarray}
where
\begin{eqnarray}
c_3& =& C_9 \sum_{Z, Z\in {\cal Z}_c} t_q \Vert h_Z \Vert
(\exp[-l_Z/\Delta E/2v']+|Z| \exp[-\mu' l_Z]).
\end{eqnarray}
Here, we use constants $c_2,c_3$ to follow the notation of \cite{suffcon}.

Eq.~(\ref{commute}) relies on the fact that $W_1(\pi)$ and $W_2(\pi)$ can
be written as exponentials of local operators ${\cal D}_1,{\cal D}_2$.  The
commutator of ${\cal D}_1$ with ${\cal D}_2$ can be bounded, and thus one
can ``re-order" the quasi-adiabatic evolution so that instead of
first evolving with $W_2$ and then with $W_1$, both flux angles are
changed at the same time, as with $W$.

Eq.~(\ref{close}) relies on the existence of a spin gap in the
system.  For a gapped system with a unique
ground state, the quasi-adiabatic evolution approximately evolves the
ground state of the initial Hamiltonian, $\Psi_0$, into the
ground state of the final Hamiltonian, $R_{\uparrow}(\pi,X)
R_{\downarrow}(-\pi,X)\Psi_0$.  Here, we have not specified that
the Hamiltonian has a gap.  However, $\partial_{\theta}
{\cal H}(\theta,-\theta,-\theta,\theta)$ is a spin-$1$ operator, and
hence the spin gap suffices to show Eq.~(\ref{close}).

We now show that $\langle W_1(\pi) \Psi_0, {\cal H}(\pi,\pi,0,0)
W_1(\pi)\Psi_0\rangle$ is close to $\langle \Psi_0, {\cal H} \Psi_0 \rangle$,
which is the desired result as by Eq.~(\ref{piispi})
it shows that $W_1(\pi)\Psi_0$ is a state
for which the expectation value of the energy with Hamiltonian
${\cal H}_{\pi}$ is close to the expectation value of the
energy of the state $\Psi_0$ with Hamiltonian ${\cal H}(0,0,0,0)$.

To show that the expectation value of the energy is close, we write
${\cal H}(\pi,\pi,0,0)=\sum_Z h_Z(\pi,\pi,0,0)$.
We define ${\cal S}_1$ to be the set of points $i$ with
$-L/4+L/4(\xi\mu+1)<x_0(i)<L/4-L/4(\xi\mu+1)$, and define 
${\cal S}_2$ to be the set of points $i$ with
$L/4+L/4(\xi\mu+1)<x_0(i)<3L/4-L/4(\xi\mu+1)$.  Note that
$W_1$ is supported on ${\cal S}_1$ and 
$W_2$ is supported on ${\cal S}_2$ as shown in Fig.~2.
If $Z\cap{\cal S}_1=\emptyset$, then 
$\langle W_1(\pi) \Psi_0, h_Z(\pi,\pi,0,0)
W_1(\pi)\Psi_0\rangle=\langle \Psi_0, h_Z \Psi_0 \rangle$.
If $Z\cap{\cal S}_2=\emptyset$, then
$|\langle W_1(\pi) \Psi_0, h_Z(\pi,\pi,0,0)
W_1(\pi)\Psi_0\rangle-\langle \Psi_0,h_Z \Psi_0 \rangle|
=
|\langle W_1(\pi) W_2(\pi) \Psi_0, h_Z(\pi,\pi,0,0)
W_1(\pi) W_2(\pi) \Psi_0\rangle-\langle 
R^{\uparrow}(\pi,X) R^{\downarrow}(-\pi,X)
\Psi_0 
, h_Z (\pi,\pi,0,0)
R^{\uparrow}(\pi,X) R^{\downarrow}(-\pi,X)
\Psi_0 \rangle|
\leq 
2(c_2+c_3)\Vert h_Z \Vert$.
Finally, the sum of $\Vert h_Z \Vert$ over
all $Z$ such that $Z\cap{\cal S}_1 \neq \emptyset$ and
$Z\cap{\cal S}_2 \neq \emptyset$ is bounded by $V s_1 \exp[-\mu L/4(\xi\mu+1)]$.
Thus, we find that
\begin{eqnarray}
&&|\langle W_1(\pi) \Psi_0, {\cal H}(\pi,\pi,0,0)
W_1(\pi)\Psi_0\rangle-\langle \Psi_0, {\cal H} \Psi_0 \rangle|\\ \nonumber
&\leq&
C_7 V^2 \sqrt{L/v \Delta E} s_1^2
\exp[-(L-L/4(\xi\mu+1))/\xi]
\\ \nonumber
&=&C_7 V^2 \sqrt{L/v \Delta E} s_1^2
(\exp[-L/4(\xi+1/\mu)]),
\end{eqnarray}
giving Eq.~(\ref{qe}) as claimed.

This calculation can be extended to systems with larger symmetry
groups than the $U(1)$ symmetry above.  If a system has $N$ species of
fermions, with an $SU(N)$ symmetry,
one may define angles $\theta_{1}^a,\theta_{2}^a$ for $a=1...N$ for each
species of fermion.  We perform a continuation
with $\theta_1^a=\theta/(N-1)$ for $1\leq a\leq
N-1$, and $\theta_1^N=-\theta$, going from $\theta=0$ to $\theta=2\pi(N-1)/N$.
Then, after the continuation we have $\theta_1^a=2\pi/N$ for
$1\leq a\leq N$, and $\theta_1^N=2\pi/N-2\pi$.  Thus, if
the ground state of the Hamiltonian with no flux
is an $SU(N)$ singlet and the system has a gap
to the lowest state which is not an $SU(N)$ singlet, then there is
a state of the Hamiltonian with flux $2\pi/N$ which is close in energy
to the ground state of the Hamiltonian without flux.  Colloquially,
the flux is quantized in units of $2\pi/N$.

\section{Discussion}
We have used quasi-adiabatic continuation as a tool to study systems
with continuous symmetries and a gap.  This leads to a version of
Goldstone's theorem for nonrelativistic systems which is valid in arbitrary
finite dimension and which does not depend on additional assumptions
regarding ergodicity.  While this result was derived from an infinitesimal
quasi-adiabatic continuation, the continuation under a non-infinitesimal
change in parameters was used to derive an approximate flux periodicity
of systems with a spin gap.  While this periodicity is realized in
different ways in different systems, for example by Byers-Yang states
in an $s$-wave superconductor and by topological excitations in
an RVB state, the mathematical result covers both cases.

Finally, we have discussed the Berry phase under quasi-adiabatic continuation,
and shown that it is close to the usual non-Abelian Berry phase for
adiabatic evolution.  This result will be used in a future work to
consider Hall conductance quantization in many-body systems, and, under
the assumption of an excitation gap, to remove the need for an averaging
assumption on the Hall conductance\cite{hc}.

{\it Acknowledgments---} I thank N. Bonesteel and T. Koma for useful
discussions.  
This work was carried out under the auspices of the NNSA of the U.S. DOE
at LANL under Contract No. DE-AC52-06NA25396.


\begin{thebibliography}{99}
\bibitem{general} V. Kalmeyer and R. B. Laughlin, Phys. Rev. Lett. {\bf 59},
2095 (1987); N. Read and S. Sachdev, Phys. Rev. Lett.
{\bf 66}, 1773 (1991); X.-G. Wen, Phys. Rev. B {\bf 44}, 2664 (1991).

\bibitem{zl} M. B. Hastings and X.-G. Wen, Phys. Rev. B {\bf 72}, 045141 (2005).

\bibitem{lsmhd} M. B. Hastings, Phys. Rev. B {\bf 69}, 104431 (2004).

\bibitem{loc} M. B. Hastings, Phys. Rev. Lett. {\bf 93}, 140402 (2004).

\bibitem{hk} M.  B. Hastings and T. Koma,
Commun. Math. Phys. {\bf 265}, 781 (2006).

\bibitem{ns} B. Nachtergaele and R. Sims, Commun. Math. Phys.
{\bf 265}, 119 (2006).

\bibitem{nbp} The idea of studying the
flux quantization problem was suggested to the author by N. Bonesteel.

\bibitem{by} N. Byers and C. N. Yang, Phys. Rev. Lett. {\bf 7}, 46 (1961).

\bibitem{e2e} S. A. Kivelson, D. S. Rokhsar, and J. P. Sethna,
Europhys. Lett. {\bf 6}, 353 (1988).

\bibitem{lr} E. H. Lieb and D. W. Robinson, Commun. Math. Phys. {\bf 28},
251 (1972).

\bibitem{la2} I. Affleck, T. Kennedy, E. H. Lieb, and H. Tasaki,
Commun. Math. Phys. {\bf 115}, 477 (1988).

\bibitem{la1} I. Affleck and E. H. Lieb, Lett. Math. Phys. {\bf 12}, 57
(1986).

\bibitem{to} T. J. Osborne, preprint quant-ph/0601019.

\bibitem{lightcone} 
S. Bravyi, M. B. Hastings, and F. Verstraete,
Phys. Rev. Lett. {\bf 97}, 050401 (2006).

\bibitem{berry} F. Wilczek and A. Zee, Phys. Rev. Lett. {\bf 52}, 2111
(1984).

\bibitem{k2} T. Koma, arxiv.org:math-ph/0505022.

\bibitem{ergodic} W. F. Wreszinski,
Fortschr. Phys. {\bf 35}, 379 (1987);
L. Landau, J. Fernando Perez and W. F. Wreszinski,
J. Stat. Phys. {\bf 26}, 755 (1981).

\bibitem{blyr} V. N. Kotov, O. Sushkov, Z. Weihong, and J. Oitmaa, Phys.
Rev. Lett. {\bf 80}, 5790 (1998).

\bibitem{blyrdefect} See Kaj H\"oglund, A. W. Sandvik, and S. Sachdev,
Phys. Rev. Lett. {\bf 98}, 87203 (2007) for a study of the behavior of a
single such defect near a quantum critical point.

\bibitem{expander} N. Alon, Combinatorica, {\bf 6(2)}, 83 (1986);
J. Friedman, Conf. Proc. of the Annual ACM
Symposium on Theory of Computing, 720 (2003); J. Friedman, preprint cs/0405020.


\bibitem{spinwave} N. W. Ashcroft and N. D. Mermin, {\it Solid
State Physics}, Chapter 33, (Harcourt Brace College Publishers, New York, 1976).

\bibitem{others} T. Senthil and M. P. A. Fisher, Phys. Rev. B {\bf 63},
134521 (2001).

\bibitem{luttinger} A. Paramekanti and A. Vishwanath, Phys.
Rev. B {\bf 70}, 245118 (2004).

\bibitem{suffcon} M. B. Hastings, Europhys. Lett. {\bf 70}, 824 (2005).

\bibitem{hc} Q. Niu and D. J. Thouless, Phys. Rev. B {\bf 35}, 2188 (1987).
\end{thebibliography}
\end{document}